**Can a composite heart rate variability biomarker shed new insights about autism spectrum disorder in school-aged children?**

Frasch, Martin G[1,2]; Shen, Chao[3]; Wu, Hau-Tieng[3,4]; Mueller, Alexander[5]; Neuhaus, Emily[6,7]; Bernier, Raphael, A.[8]; Kamara, Dana[9]; and Beauchaine, Theodore, P.[9]

[1] Department of Obstetrics and Gynecology, University of Washington, Seattle, WA
[2] Center on Human Development and Disability, University of Washington, Seattle, WA
[3] Department of Mathematics, Duke University, Durham, NC
[4] Department of Statistical Science, Duke University, Durham, NC
[5] Innere Medizin 1, Department of Cardiology, Klinikum rechts der Isar, Technical University of Munich, Germany
[6] Seattle Children's Research Institute, Center for Child Health, Behavior, and Development, Seattle, WA
[7] Seattle Children's Autism Center, Seattle, WA
[8] Department of Psychiatry and Behavioral Sciences, University of Washington, Seattle, WA
[9] Department of Psychology, The Ohio State University, Columbus, OH

## Abstract

High frequency heart rate variability (HRV) has identified parasympathetic nervous system alterations in autism spectrum disorder (ASD). In a cohort of school-aged children with and without ASD, we test a set of alternative linear and nonlinear HRV measures, including phase rectified signal averaging, applied to a segment of resting ECG, for associations with ASD vs. other psychiatric conditions. Using machine learning, we identify HRV measures derived from time, frequency, and geometric signal-analytical domains that (1) identify children with ASD relative to peers with receiver operating curve area of .89, and (2) differentiate such children from those with conduct problems or depression. Despite the small cohort and lack of prospective external validation, these preliminary results warrant larger prospective validation studies.

**Key Words:** heart rate variability; electrocardiogram; biomarker

**Corresponding author**
Emily Neuhaus, PhD
Seattle Children's Research Institute
Center for Child Health, Behavior, and Development
Seattle, WA
Email: eneuhaus@uw.edu

**Abstract**

Several studies show altered heart rate variability (HRV) in autism spectrum disorder (ASD), but findings are neither universal nor specific to ASD. We apply a set of linear and nonlinear HRV measures—including phase rectified signal averaging—to segments of resting ECG data collected from school-age children with ASD, age-matched typically developing controls, and children with other psychiatric conditions characterized by altered HRV (conduct disorder, depression). We use machine learning to identify time, frequency, and geometric signal-analytical domains that are specific to ASD (receiver operating curve area = .89). This is the first study to differentiate children with ASD from other disorders characterized by altered ASD. Despite a small cohort and lack of external validation, results warrant larger prospective studies.

## Introduction

Autism spectrum disorder (ASD) is a neurodevelopmental disorder characterized by significant social, communication, and behavioral challenges. ASD affects ~1 in 60 children, according to Centers for Disease Control estimates (CDC, 2018), and appears across all racial, ethnic, and socioeconomic groups (e.g., Neuhaus, Beauchaine, Bernier, & Webb, 2018). Substantial research indicates that genetic influences underlie ASD (Arnett et al, 2018; Coe et al, 2019), with subsequent differences in biomarkers observed across a range of physiological systems (e.g., brain structure and function, neurotransmitter systems) and research modalities (e.g., neuroimaging, electrophysiology) (Li et al, 2017; Zwaigenbaum & Penner, 2018). As a field, identification of biomarkers that correspond with ASD contributes to broad goals of better understanding etiology, improving diagnosis and screening practices, and assessing change over time and treatment (Walsh, Elsabbagh, Bolton, & Singh, 2011).

Within the array of indices identified thus far, heart rate variability (HRV) emerges as a promising biomarker of ASD—particularly with regard to social difficulties. HRV signal is quantified via a number of different metrics with the goal of extracting physiologically meaningful biomarkers. HRV measures attributed to the parasympathetic nervous system's (PNS) influence on the electrocardiogram (ECG) include respiratory sinus arrhythmia (RSA), root mean successive square difference of R-R intervals of ECG (RMSSD), and high frequency (HF; usually in the range of $0.15 - 2$ Hz but varying from study to study) spectral power (see e.g., Beauchaine, 2001; Benevides & Lane, 2015; Porges, 2007). Within the autonomic nervous system, PNS, among many things, reduces heart rate and facilitates flexible, adaptive engagement with the environment. The vagus nerve, the longest nerve connecting brain and body, represents the major anatomical and physiological substrate of the PNS. As described by several theoretical frameworks (e.g., Polyvagal Theory, Porges, 2003; Neurovisceral Integration

Theory, Thayer & Lane, 2000), relatively higher vagal activity, reflected in changes of HRV patterns, should enable more adaptive responses to social demands.

Consistent with this, a large literature across typically developing and clinical samples links HRV with flexible and effective social engagement throughout childhood and adolescence (see e.g., Beauchaine, 2001; Beauchaine & Thayer, 2015; Porges, 2007). Higher baseline (resting) HRV correlates with greater social competence in a number of forms, including stronger social responsiveness, more frequent use of eye gaze during social interaction, greater degree of sympathy toward others, and more frequent expressions of concern for others in distress, (Doussard-Roosevelt, Porges, Scanlon, Alemi, & Scanlon, 1997; Fabes, Eisenberg, & Eisenbud, 1993; Heilman, Bal, Bazhenova, & Porges, 2007; Patriquin, Lorenzi, Scarpa, & Bell, 2014; Taylor, Eisenberg, & Spinrad, 2015). Robust associations between PNS function and prosocial behavior among typically developing groups suggests the possibility of altered HRV among those with social deficits, including those with autism.

Existing research highlights possible reductions in HF HRV, a linear frequency domain measure of HRV, as well as in RSA, among individuals with ASD, but findings are not universal (Benevides & Lane, 2015; Billeci et al., 2018). Numerous studies comparing resting HRV (mostly assessed via RSA) among children, adolescents, and adults with ASD versus typically developing peers document lower resting HRV (Bal et al., 2010; Bujnakova et al., 2016; Ming, Julu, Brimacombe, Connor, & Daniels, 2005; Guy, Souders, Bradstreet, DeLussey, & Herrington, 2014; Thapa et al., 2019; Vaughan Van Hecke et al., 2009). Developmentally, whereas children without ASD demonstrate increasing HRV measured with RSA throughout infancy and early childhood (see Shader et al., 2018), those later diagnosed with ASD display a slower rate of increase in RSA and lower resting RSA by age 18 months relative to peers (Sheinkopf et al 2019). Thus, longitudinal data suggest that the developmental trajectory of HRV may begin to diverge as early as infancy among children later diagnosed with ASD. Of note, so far longitudinal development of HRV among children with ASD has been quantified exclusively by RSA, and not by any other linear or nonlinear measures.

In contrast to these findings are studies in which resting measures of HRV (SDNN, RMSSD, HF, LF, LF/HF ratio, RSA) do not differ between children with ASD versus those without (Daluwatte et al., 2012; Levine et al., 2012; Sheinkopf, Neal-Beevers, Levine, Miller-Loncar, & Lester, 2013; Watson, Roberts, Baranek, Mandulak, & Dalton, 2012). Methodological approaches have varied substantially across the field thus far, and differences related to stimulus conditions, task designs, data processing, participant characteristics, and similar factors may account in part for variation in findings in this literature (Benevides & Lane, 2015; Patriquin et al., 2019). As a result of heterogeneities in research approaches and findings, ASD-related differences in PNS function require further study.

These concerns are further complicated by the fact that alterations in HRV and PNS function, as measured and analyzed to date in the literature, are not unique to ASD. Instead, reduced baseline HRV, as quantified by the above mentioned HRV measures, characterizes a number of psychiatric disorders and high-risk traits, including depression, anxiety, non-suicidal self-injury, and conduct problems (Beauchaine, 2001; Crowell et al., 2005; Hastings et al., 2008; Koenig, Kemp, Beauchaine, Thayer, & Kaess, 2016). Despite this overlap, studies of HRV in ASD have not routinely included clinical comparison groups. Doing so may yield a more fine-grained understanding of HRV in ASD, as inclusion of participants with broader developmental and/or psychiatric concerns may allow researchers to delineate points of divergence that are specific to ASD.

The research on HRV and deploying the various HRV measures to predict a wealth of health outcomes have now entered its 5th decade, yet the conceptualization of the physiological origins of HRV and the meaning of its complex behavioral response patterns remain a work in progress. Consequently, any selection of particular HRV measures for a research question will remain somewhat subjective. From the perspective of HRV methodologies, the focus in studies of ASD and other neurodevelopmental disorders has been on linear HRV measures mentioned above. There are no studies in this field that have considered the plethora of the nonlinear HRV characteristics that have been developed, tested in preclinical and clinical studies and shown to perform well in different clinical scenarios in identifying pathophysiological states (Bravi, Longtin, & Seely, 2011; Herry et al., 2019; Sassi et al., 2015; Shaffer & Ginsberg, 2017). HRV, assessed in different signal-analytical domains at the same time, represents a complex pattern that can be decoded to predict physiological responses such as the response to an inflammatory stimulus (Durosier et al., 2015; Herry et al., 2016), lipopolysaccharide (LPS), hypoxia (Frasch et al., 2014), or a homeokinetic response to surgery stress (Herry et al., 2019). Due to the fundamentally nonlinear nature of HRV, a complete vagal withdrawal by cervical bilateral vagatomy causes a shift in the entire landscape of HRV measures, rather than a change in a single measure (Herry et al., 2019). These shifts in HRV, occurring due to inflammation, hypoxia, surgery or vagotomy, underscore the complex nature of HRV which cannot be captured by any single measure and, therefore, cannot be quantified as increasing or decreasing per se (Frasch et al., 2018). Overall, an opportunity exists to expand the scope of known changes in HRV due to ASD by considering additional, nonlinear aspects of HRV. The HRV measures we selected for this study all have in common that they have been tested to reflect integrative physiological behavior involving autonomic nervous system (ANS) and the vagal function in particular. This approach may lead to novel biomarkers of ASD and its development.

We aim to better characterize HRV among children with ASD by considering, in tandem, two datasets collected within the same laboratory for related but distinct research purposes. In a novel reanalysis, we consider whether a composite HRV measure, combined with machine learning, can yield information about PNS function in ASD versus other psychiatric disorders. By capitalizing on existing datasets collected from children in four diagnostic categories (ASD, conduct problems, depression, typical development), we evaluate specificity of a composite HRV index with regard to ASD. Thus, we seek to better characterize specific associations between HRV and ASD by applying a novel set of linear and nonlinear HRV measures.

## Method

### Procedure

We performed secondary analyses of HRV data described elsewhere [author citation]. Data were collected originally as part of two IRB-approved studies at [author citation]. Data from four groups of participants were extracted: ASD, conduct problems (CP), depression, and typical development (TD). Within each study, potential eligibility for participation was ascertained through phone interviews with caregivers. During subsequent lab visits, informed consent and assent were obtained and diagnoses were verified through fact-to-face interviews with parents. For all groups, baseline cardiac data were collected over the course of a 5-min rest period. This was followed by experimental tasks that are not pertinent to this study [author citation].

### Participants

Data from 69 participants (18 with ASD, 18 with CP, 15 with depression, 18 TD) were included. All participants were between the ages of 8 and 12 years and were fluent English speakers. The overall racial/ethnic distribution was as follows: 12.1% African American, 7.6% Asian/Pacific Islander, 63.6% Caucasian, 3.0% Latino, 3.0% other, and 10.6% more than one race. Distribution of race/ethnicity did not differ across the four diagnostic groups according to conventional statistical standards, $\chi_2(15)=20.44$, $p=.16$. However, post hoc tests suggested that participants who identified as African American were overrepresented in the depression group, and those who endorsed "more than one race" were overrepresented in the TD group.

Inclusion criteria differed by group. Children with ASD (18 male, 0 female; mean age=119.8 months, *SD*=13.2) met CPEA (Collaborative Programs of Excellence in Autism; Lainhart et al., 2006) cut-off criteria on the Autism Diagnosis Observation Schedule (ADOS; Lord, Rutter, DiLavore, & Risi, 2003) and Autism Diagnostic Interview-Revised (ADI-R; Lord, Rutter, & Le Couteur, 1994). Many entered the study with an existing ASD diagnosis from a community provider. Previous diagnoses were confirmed and first-time diagnoses were determined by an experienced clinician according to the *DSM-IV* (American Psychiatric Association, 1994) criteria for autism, Asperger's, or PDD-NOS using all available information.

Children in the CP group (13 male, 2 female; mean age=123.0 months, *SD*=17.90) met *DSM-IV* criteria for CD and/or oppositional defiant disorder on the parent-report Child Symptom Inventory (CSI; Gadow & Sprafkin, 1997). Sensitivity and specificity of the CSI are adequate to excellent, and demonstrate strong convergence with comprehensive diagnostic tools (Sprafkin, Gadow, Salisbury, Schneider, & Loney, 2002). Children in the depression group (10 male, 5 female; mean age=126.0 months; *SD*=18.23) met criteria for either major depression or dysthymia, based on corresponding CSI subscales. For participants included herein, comorbidity for CD and depression was a rule-out.

Children in the TD group (18 male, 0 female; mean age=120.1 months, *SD*=11.10) were included based on absence of any personal history of psychiatric diagnoses, seizures, or head injury; family history of ASD; a total score exceeding 9 on the lifetime version of the Social Communication Questionnaire (SCQ; Rutter, Bailey, & Lord, 2003); and a *T*-score > 65 on the thought problems scale of the Child Behavior Checklist (CBCL; Achenbach & Edelbrock, 1991).

**Measures and Tasks**

*Social behavior among the ASD and TD groups.* Parents of children in the ASD and TD groups completed several measures of children's social function. First, the social problems scale *T*-score from the CBCL was computed as a measure of social difficulties. Parents also completed the Social Skills Improvement System (SSIS; Gresham & Elliott, 2008), from which the Social Skills standard score was obtained. Finally, parents were interviewed with the Vineland Adaptive Behavior Scale, 2nd Ed. (Sparrow, Cicchetti, & Balla, 2005). Standard scores from the Socialization domain were obtained.

*Data acquisition.* Electrodes used to collect electrocardiographic (ECG) data were applied in a modified Lead II configuration. Following electrode placement, participants sat upright in an appropriately-sized chair for 5 min while baseline data were collected. During this baseline, participants were instructed to sit quietly, were given no additional task demands or activities, and were unaccompanied by parents or research personnel. The data acquisition room was sound-attenuated room to minimize ambient noise and furnished simply in a developmentally appropriate style to increase participant comfort and minimize distraction. An external ECG signal was obtained using a Grass Model 15LT Physiodata Amplifier System (West Warwick, RI), sampled and digitized for later scoring at 1 kHz. ECG data were collected using COP-WIN software, version 6.10 (Bio-Impedance Technologies, Chapel Hill, NC). All participants tolerated collection of autonomic data. Data from one participant with ASD were removed from HRV analysis due to signal artifacts. No participant's data were removed from the other cohorts.

**ECG and R-R Times Series Pre-processing**

R-R interval time series preprocessing is a prerequisite for estimating HRV accurately. R-R intervals were inspected for possible artifacts (e.g., missed or spurious heartbeats). We assumed that participants were free of non-respiratory-related arrhythmias. Epochs in which numbers of apparent artifacts exceeded a priori cut-offs were discarded (see Table S1). R-R intervals that were longer than 2000 ms (suggesting HR slower than 30 bpm) or shorter than 300 ms (suggesting HR faster than 200 bpm) were removed and interpolated. We considered these settings appropriate for resting baseline recordings (Peltola, 2012). There were no group differences in the number of artifacts removed. Following preprocessing, there were 18 children in the TD cohort, 17 children in ASD cohort, 18 children in CP cohort, and 15 children in depression cohort. Thus, we used R-R interval time series from $N$=68 participants to compute HRV measures and build classification models.

The detailed R-R intervals adjustment methodology is as follows:

**Step 1:** for each R-R interval less than threshold-low, merge it to the RRI ahead of it;

**Step 2:** for each bad R-R interval greater than threshold-high, collect first 30 neighbor R-R intervals, then calculate their median; then, the bad R-R interval is divided into smaller R-R intervals of equal length, such that their lengths are closest to the median;

**Step 3:** remove R-R interval spikes by interquartile range (IQR), that is,

for each R-R interval, first collect 30 neighbor R-R intervals, then the **IQR = 75% quantile - 25% quantile**;

if this R-R interval is less than **25% quantile - 1.5\*iqr :** do the same as in step 1;

if this R-R interval is greater than **75% quantile + 1.5\*iqr :** do the same as in step 2.

**HRV**

We selected a set of linear and nonlinear HRV measures for analysis (see Table S2), including standard deviation of NN intervals (SDNN), root mean square of the successive differences (RMSSD), low frequency (LF) spectral power, high frequency (HF) spectral power, sample entropy, fuzzy entropy, skewness, asymmetry index, Poincaré plot indices, recurrence quantification analysis indices, and phase rectified signal averaging (PRSA). Selection of HRV measures was based on four complementary considerations, including:

(1) HRV Task Force recommendations comprising linear HRV measures in time and frequency domain,

(2) Previous work in which several measures perform well in R-R interval time series analyses to separate experimental and control groups (e.g., Beauchaine, 2001, 2012; Berry, Palmer, Distefano, & Masten, 2019),

(3) Addition of complementary HRV measures from signal-analytical domains yet unexplored in ASD domain (complexity, geometric, statistical, PRSA) to characterize changes in ASD more comprehensively than before and

(4) Proven performance of the nonlinear HRV measures as integrative biomarkers of inflammation, hypoxia, cardiac performance and stress chosen from the complementary signal-analytical domains (Table S2; Frasch et al., 2018). These nonlinear biomarkers are used in ASD studies for the first time.

Time domain metrics of HRV included SDNN and RMSSD. For frequency domain metrics, respiratory bands used to define LF and HF spectral power were $< 0.12$ Hz and $> 0.12$ Hz, respectively, consistent with previous research conducted by our group and others. We used fast-Fourier analysis of the 5-min R-R time series to extract spectral power. Notably, 5 min is shorter than ideal for estimating LF power (Task Force, 1996). Nevertheless, we included LF power given its popularity.

The remaining HRV measures belong to several signal-analytic domains (Herry et al., 2016). Adding these is a strength of the present approach, as it provides complementary information and a more comprehensive capture of physiological characteristics embedded in HRV than possible with any given single measure. We demonstrate performance of complementary HRV measures from different signal-analytical domains.(Herry et al., 2016) We cite some known precedents for performance of the chosen HRV measures under physiological conditions next to each measure (Table S2).

**Phase-rectified Signal Averaging (PRSA)**

An R-R interval time series contains oscillations of different frequencies, each frequency reflecting a physiological regulatory process controlled by an intrinsic closed-loop control. Magnitudes of oscillations provide insight into functional states of the cardiovascular system and provide prognostic information in certain contexts. HRV analysis with PRSA provides for detailed consideration of well-defined physiological processes such as HR acceleration or deceleration (Bauer, Kantelhardt, Barthel, et al., 2006; Bauer, Kantelhardt, Bunde, et al., 2006).

The basic principle of PRSA is to align segments of R-R interval time series to a predefined feature, which we call an anchor. In the case of heart rate analysis, this is typically a beat-to-beat change in R-R intervals. The procedure consists of three following steps. In the first, R-R intervals are divided into two categories: the so-called deceleration anchor (defined as intervals that are longer than the previous interval) and the acceleration anchor (defined as intervals that are shorter than the previous interval). Figure 1 (Panel 1) shows a short part of a R-R interval time series with anchor points for decelerations (Panel A) and accelerations (Panel B) marked.

To exclude too large beat-to-beat differences (e.g., artifacts), a maximum allowed change (FMax) can be defined. In the case of R-R interval prolongation, this can, for example, be a maximum of 105% of the previous interval. Thus, only extensions between two R-R intervals are used as anchors, which have a prolongation of 105%. A minimum necessary change (FMin) can also be defined. Using the example of R-R interval prolongation, an FMin=102% means that the interval must be at least 102% longer than the previous interval. For anchor points to be determined, FMax must always be greater than FMin.

Setting FMin and FMax allows for definition of direction and minimum and maximum change of beat-to-beat differences. The anchor criterion FiltLow=90 % and FiltHigh<100 % only determines shortenings of the R-R intervals (i.e., HR accelerations as anchor points). In order to only consider HR decelerations, FiltLow must be selected at least >100% and FiltMax >= 101%. A combination of HR accelerations and HR decelerations as anchor definition can be achieved by FMin<100% and FMax>100%. Definition of segments in sample R-R interval time series is visualized in the top part of Figure 1 (Panel 2). Here, only windows for deceleration-related anchor points are shown.

In the second step of PRSA, windows of length 2L around each anchor point are defined with anchor points at the center of windows. As shown in Figure 1 (Panel 2), selected segments can overlap. The number of these windows corresponds to the number of anchor points found (M).

In the third step, a phase-rectified average (PRA) signal is calculated by first aligning all windows with their anchor points at the center (Figure 1, Panel 2, middle part). Then, R-R interval values are averaged over all M windows. Aligned windows build a complex picture comprising all detected oscillations around each anchor point (Figure 1, Panel 2, lower part). Visual inspection of these windows leads to no conclusion about captured oscillations. Only after averaging do oscillations become clearly apparent.

In our example, the PRA signal contains one relevant periodic oscillation, which can be seen clearly. Note that PRSA allows for assessment of the periodic content of long-term recordings by examining a rather short signal; all important periodic or quasiperiodic components of the original R-R interval time series are centered on defined anchor points.

To quantify information from the PRA signal we take one specific coefficient of the Haar wavelet analysis (Eq. 1). Here we focus the central part of the PRA signal which reflects the magnitude of all oscillations which are here in phase (Figure 1, Panel 3).

$$PRSA_{Magnitude} = \frac{1}{4}\left(PRA_{(0)} + PRA_{(1)} - PRA_{(-1)} - PRA_{(-2)}\right) \qquad \text{(Equation 1)}$$

In summary, PRSA quantifies the mean amplitude of all oscillations of heart rate. The time series of R-R intervals is a composite output of different autonomic regulations (see e.g., Beauchaine, 2001, Fig. 1, p. 187). Notably, PRSA can extract periodic components out of non-stationary signals (Bauer, Kantelhardt, Barthel, et al., 2006; Bauer, Kantelhardt, Bunde, et al., 2006). This contrasts with spectral analyses, which assumes stationarity.

We sought to identify optimal filter settings for anchor point definition, so we varied lower and upper filter settings. Lower and upper filters build the range of interval length that is accepted as anchors. We varied both filters from 60% to 140% in steps of 2%. For each valid combination of lower and upper filter settings, we calculated the PRSA values for all measures. We then calculated areas under receiver operating curve (AUCs) to determine diagnostic power. The maximum AUC of .76 was found using a lower filter of 90% and an upper filter of 128%. This means we have to combine information about heart rate decelerations and accelerations to maximize group differences.

**Participant Classification Using Machine Learning**

Next, we specified a set of machine learning models and identified the best performer for correct classification of participants into the four groups (DataIKU, New York City, NY). We used AUC ROC as a scoring performance criterion. Best performance was achieved by XGBoost approach (Chen et al., 2016). XGBoost stands for "Extreme Gradient Boosting", where the term "Gradient Boosting" originates from (Friedman, 2001). XGBoost is a classification algorithm, optimized for speed and performance. We ran a 5-fold cross validation; that is, we used a standard 80:20 training:validation split to train the model and test model performance. See GitHub for code details and to reproduce the approach.

**Statistical Approach**

We assumed significant difference in psychometric scores and HRV measure values at $p <$ .05. AUCs were calculated for each HRV measure to express potential for distinguishing children with ASD from others.

## Results

Table 1 reports clinical characteristics of the ASD and TD cohorts (author citation). As previously reported (author citation), we found SSIS and VABS differed at 79.8±10.7 and 77.6±17.7 in children with ASD compared to 101.3±10.8 and 119.3±6.9, respectively, in children without ASD ($p < .01$). Among children with ASD, ADI-R total score averaged 41.8±8.5 and ADOS severity score averaged 7.8±1.6.

HRV measures for each group are reported in Table 2 as a reference set. We deliberately avoided performing individual group comparisons for each HRV measure, because our aim was to determine if HRV measures *in toto* could distinguish the ASD group from the other groups. Accordingly, Table 3 shows performance results of all machine learning models we evaluated. With regard to the ASD cohort, XGBoost model performed best not only on ROC AUC, but also on precision and recall metrics, rendering ROC AUC of .88. Figure 2 subplot "HRV measure importance" ranks the importance of the HRV measures studied in their contribution to the XGBoost model performance. The top five HRV measures that contributed 77% of the model performance were the mode skewness and SD2 (23% and 17%, respectively), followed by PRSA, RMSSD, and moment coefficient of skewness (measures of asymmetry) (Figure 2).

**Discussion**

Parasympathetic markers including those derived from HRV are clearly associated with social function in the context of typical development (Doussard-Roosevelt et al., 1997; Fabes et al., 1993; Heilman et al., 2007; Patriquin et al., 2014; Taylor et al., 2015), prompting increasing interest in their relations to ASD. Through analyses described herein, we present a hypothesis-generating retrospective study aimed at discovering novel HRV measures that are useful in characterizing ANS patterns associated with ASD. As described above, numerous studies indicate reduced magnitude of HRV measures such as RSA and HF spectral power in ASD relative to controls, but findings are not universal (see Benevides & Lane, 2015). Our approach combines multiple HRV measures and machine learning to identify a composite HRV metric with very good identification of ASD children (ROC AUC = .88) within a mixed set of age-matched participants with typical development, CP, or depression. Noteworthy, the latter two conditions are also identified with high ROC AUC values. Pending external validation, our results identify a set of eleven HRV measures including asymmetry, Poincaré plot, recurrence quantification analysis and entropy indices as well as PRSA. We present a corresponding machine learning model capable of identifying children with ASD prospectively.

Because our sample included children with other clinical concerns, our findings also provide some evidence that this composite HRV approach may be helpful in identifying features that have greater specificity to ASD. To date, studies of HRV in children with ASD have not routinely included clinical comparison groups despite findings of altered ANS function in those groups. Doing so may allow for clearer understanding of how ANS function relates to core and associated deficits of ASD and other diagnoses. Although HRV alterations may be a transdiagnostic feature of a variety of clinical concerns (e.g., Beauchaine, 2015), it may also be that approaches such as ours can identify particular sets of HRV biomarkers that carry specificity to ASD. This remains an empirical question that has not yet been fully explored.

Our approach to selecting the HRV measures was based on a body of evidence from the HRV literature across signal-analytical domains. Across various contexts, these HRV measures reflect changes in ANS activity due to inflammation, hypoxia, and cardiac intrinsic dynamics (Frasch et al. 2009, 2017, 2018; Herry et al., 2016). Many of these HRV measures are not commonly tested, in part because they are not included in standard HRV software packages.

In a recent study (Herry et al., 2019), we identified a set of measures as comprising a generic HRV code not specific to vagal modulations of HRV, as well as a set of four HRV measures specific to vagal modulation of HRV (so-called vagus code, cf. Fig. 4 in Herry et al., 2019). Among measures of the subset reflecting a generic HRV code, we identified Fuzzy Entropy and Poincaré plot SD2. The latter is mathematically equivalent to RMSSD in a different signal-analytical domain, also identified here as associated with ASD. The vagus code subset includes asymmetry index quantifications of which we also found in the present study to be among the most important HRV measures (skewness indices) differentiating the ASD cohort. Overall, these measures are derived from time, frequency, geometric, complexity and statistical signal-analytical domains and together comprise a predictor of changes in HRV reflective of ASD. Given appropriate stimulus conditions, RMSSD also reflects vagal influences on cardiac output, as determined by pharmacologic blockade (e.g., Task Force, 1996).

**Limitations and Next Steps**

Our results must be interpreted in the context of sample characteristics, particularly with regard to our ASD group. For both our ASD and TD groups, for whom standardized cognitive tests were completed, cognitive skills were relatively high and all children were verbally fluent. Given the heterogeneity ASD, our findings may be most generalizable to a subset of individuals who have comparable cognitive and verbal abilities, and less generalizable to those with different cognitive and verbal profiles. Similarly, all participants tolerated testing procedures, including sensory demands of wearing adhesive sensors, behavioral demands of remaining sufficiently still during data collection, and verbal demands of understanding study instructions. Thus, the sample likely best represents children with similar abilities, and our findings may not extend fully to children with greater sensory, behavioral, or communication difficulties. That said, the "high functioning" nature of our ASD group may serve to increase their similarity to children in the CP, depression, and TD groups, thus yielding a more stringent test of HRV measures intended to differentiate between them.

Caveats aside, our key findings include (1) identification of possible HRV "signatures" for children with ASD at baseline; (2) derivation of a machine learning model based on HRV data that distinguishes among children with ASD, typical development, and both CP and depression.

Our findings now warrant external, prospective validation in larger samples of children and adolescents. We are optimistic that such validation studies will help further refine a composite set of qualitative and quantitative HRV measures with greater specificity for behavioral correlates identifying individuals on the autism spectrum. As the presented approach is refined through such external validation studies, it is possible that such composite HRV measures can aid in screening and diagnostic efforts among children with symptoms of ASD, contribute to models of etiology and pathophysiology of ASD, and provide biomarkers with which to track development and response to intervention among children with ASD in an accessible, noninvasive, and inexpensive manner in both research and clinical settings.

## Compliance with Ethical Standards

**Ethical approval**: All procedures performed in studies involving human participants were in accordance with the ethical standards of the institutional and/or national research committee and with the 1964 Helsinki declaration and its later amendments or comparable ethical standards.

**Informed consent**:  Informed consent was obtained from all individual participants included in the study.

**Figure Legends**

*Figure 1.* **Phase rectified signal averaging (PRSA) approach.**

**Panel 1.** Anchor points related to prolongations between consecutive RR intervals (green dots). (B) Anchor points related to shortenings between consecutive RR intervals (red dots).

**Panel 2.** The upper part shows a part of the heartbeat interval time series; light gray dots mark deceleration-related anchor points. Around each anchor point, windows are defined with length 2L. Here, L was chosen to be 10 heartbeat intervals before and 10 heartbeat intervals after corresponding anchor points. For phase rectification, all windows are aligned with anchor points at the center (middle part). Averaging over all windows results in the PRA signal (lower part).

**Panel 3.** Intervals of deceleration-related PRSA signal used for calculation of PRSA-magnitude (as shown in Eq. 1).

*Figure 2.* Classification performance of HRV measures in identifying children with ASD among participants considered as showing typical development (TD) or with conduct problems (CP) or depression.

**Table 1.**

**Means and Standard Deviations Behavioral Variables for ASD and TD Groups**

| | Control | (*SD*) | ASD | (SD) | *F* | Effect size (*d*) |
|---|---|---|---|---|---|---|
| **Full scale IQ** | 114.8 | 13.5 | 108.3 | 21.4 | 1.17 | 0.03 |
| **ADI-R total score** | -- | -- | 41.8 | 8.5 | -- | -- |
| **ADOS Calibrated Severity Score** | -- | -- | 7.8 | 1.6 | -- | -- |
| **SCQ total score** | 3.0 | 2.3 | 19.7 | 5.0 | 172.65*** | 0.84 |
| **SSIS social skills** | 101.3 | 10.8 | 79.8 | 10.7 | 36.13*** | 0.52 |
| **CBCL social problems** | 53.8 | 4.5 | 62.6 | 7.6 | 17.90*** | 0.35 |
| **Vineland-2 socialization** | 119.3 | 6.9 | 77.6 | 17.7 | 87.18*** | 0.72 |

*Notes*. SCQ=Social Communication Questionnaire; SSIS=Social Skills Improvement System; CBCL=Child

Behavior Checklist; Vineland–2=Vineland Adaptive Behavior Scales (2nd Ed.).

***$p<.001$.

**Table 2.**

**Means and Standard Deviations of Heart Rate Variability Measures by Group**

| | SDNN | (*SD*) | RMSSD | (*SD*) | LF power | (*SD*) | HF power | (*SD*) | Poincare SD1 | (*SD*) | SD2 | (*SD*) | Entropy SampEn | (*SD*) | FuzzEn | (*SD*) |
|---|---|---|---|---|---|---|---|---|---|---|---|---|---|---|---|---|
| **TD** | 0.80 | 0.29 | 0.82 | 0.37 | 0.003 | 0.001 | 0.006 | 0.003 | 0.59 | 0.27 | 0.94 | 0.34 | 0.95 | 0.38 | 0.16 | 0.03 |
| **ASD** | 0.60 | 0.23 | 0.56 | 0.30 | 0.002 | 0.001 | 0.004 | 0.002 | 0.40 | 0.22 | 0.73 | 0.26 | 0.88 | 0.26 | 0.14 | 0.04 |
| **CD** | 0.08 | 0.04 | 0.08 | 0.06 | 0.001 | 0.001 | 0.002 | 0.001 | 0.06 | 0.05 | 0.09 | 0.04 | 1.16 | 0.45 | 0.14 | 0.04 |
| **Depression** | 0.07 | 0.02 | 0.07 | 0.04 | 0.000 | 0.000 | 0.002 | 0.001 | 0.05 | 0.03 | 0.08 | 0.03 | 0.98 | 0.24 | 0.14 | 0.03 |

| | Asymmetry Index MC* | (*SD*) | Mode** | (*SD*) | Median*** | (*SD*) | Recurrence Quantification Analysis RR | (*SD*) | DET | (*SD*) | ENTR | (*SD*) | L | (*SD*) | PRSA | (*SD*) |
|---|---|---|---|---|---|---|---|---|---|---|---|---|---|---|---|---|
| **TD** | 5.16 | 10.22 | 0.35 | 0.45 | 0.29 | 0.32 | 0.06 | 0.03 | 0.69 | 0.07 | 0.95 | 0.19 | 3.01 | 0.51 | 6.08 | 4.98 |
| **ASD** | 7.68 | 14.98 | 0.23 | 0.42 | 0.11 | 0.37 | 0.08 | 0.09 | 0.71 | 0.09 | 1.03 | 0.31 | 3.20 | 0.95 | 2.69 | 3.37 |
| **CD** | 0.36 | 0.70 | 0.60 | 0.75 | 0.17 | 0.41 | 0.07 | 0.03 | 0.72 | 0.08 | 1.03 | 0.20 | 3.15 | 0.43 | 4.07 | 3.42 |
| **Depression** | 0.68 | 0.69 | 0.08 | 0.79 | 0.30 | 0.35 | 0.07 | 0.02 | 0.74 | 0.08 | 1.07 | 0.18 | 3.27 | 0.36 | 4.41 | 3.31 |

* MC, moment coefficient of skewness

** Mode, mode skewness

*** Median, median skewness

**Table 3.**

**Performance of Machine Learning Models Using HRV Measures for Group Classification***

| ML Model | ROC MAUC** | TD | ASD | CP | Depression | Accuracy | Precision | Recall |
|---|---|---|---|---|---|---|---|---|
| *XGBoost* | *0.85 (± 0.12)* | 0.911 | ***0.879*** | 0.975 | 0.958 | *0.59 (± 0.22)* | *0.57 (± 0.22)* | *0.59 (± 0.14)* |
| Extra trees | 0.84 (± 0.03) | 0.867 | 0.848 | 0.825 | 0.833 | 0.56 (± 0.15) | 0.53 (± 0.17) | 0.55 (± 0.19) |
| K Nearest Neighbors (k=5) | 0.83 (± 0.11) | 0.9 | 0.742 | 0.85 | 0.813 | 0.47 (± 0.22) | 0.49 (± 0.21) | 0.50 (± 0.25) |
| Random forest | 0.83 (± 0.06) | 0.889 | 0.788 | 0.825 | 0.833 | 0.50 (± 0.08) | 0.48 (± 0.14) | 0.50 (± 0.15) |
| Artificial Neural Network | 0.82 (± 0.16) | 0.933 | 0.939 | 1 | 0.958 | 0.44 (± 0.33) | 0.40 (± 0.39) | 0.47 (± 0.38) |
| Logistic Regression | 0.82 (± 0.08) | 0.822 | 0.576 | 1 | 1 | 0.54 (± 0.14) | 0.53 (± 0.20) | 0.55 (± 0.19) |
| SVM | 0.81 (± 0.06) | 0.956 | 0.545 | 0.9 | 0.875 | 0.47 (± 0.09) | 0.44 (± 0.09) | 0.49 (± 0.08) |
| SGD | 0.80 (± 0.19) | 0.878 | 0.47 | 0.95 | 1 | 0.44 (± 0.26) | 0.47 (± 0.36) | 0.48 (± 0.28) |
| LASSO-LARS | 0.79 (± 0.11) | 0.933 | 0.545 | 0.95 | 0.917 | 0.47 (± 0.26) | 0.44 (± 0.35) | 0.47 (± 0.30) |
| Gradient Boosted Trees | 0.75 (± 0.17) | 0.889 | 0.848 | 0.925 | 0.958 | 0.50 (± 0.19) | 0.48 (± 0.27) | 0.49 (± 0.26) |
| Decision Tree | 0.71 (± 0.22) | 0.489 | 0.394 | 0.688 | 0.646 | 0.52 (± 0.33) | 0.53 (± 0.38) | 0.54 (± 0.35) |

* TD, typical development; CP, conduct problems

** ROC MAUC, Receiver Operating Curve Multi-Area Under the Curve

**Table S1.**

**Numbers of R-R Interval Artifacts Removed by Participant and Group**

| Beats removed* | Group |
|---|---|
| 1 | TD |
| 0 | TD |
| 0 | TD |
| 0 | TD |
| 5 | TD |
| 1 | TD |
| 1 | TD |
| 1 | TD |
| 0 | TD |
| 0 | TD |
| 0 | TD |
| 0 | TD |
| 0 | TD |
| 0 | TD |
| 0 | TD |
| 0 | TD |
| 0 | TD |
| 0 | TD |
| 0 | ASD |
| 0 | ASD |
| 0 | ASD |
| 0 | ASD |
| 0 | ASD |
| 0 | ASD |
| 2 | ASD |
| 8 | ASD |
| 0 | ASD |
| 1 | ASD |
| 0 | ASD |
| 0 | ASD |
| 0 | ASD |
| 0 | ASD |
| 1 | ASD |
| 0 | ASD |
| 0 | ASD |
| 11 | ASD |
| 4 | CP |
| 2 | CP |
| 7 | CP |
| 0 | CP |
| 0 | CP |
| 1 | CP |
| 0 | CP |
| 0 | CP |
| 0 | CP |
| 3 | CP |
| 7 | CP |
| 6 | CP |
| 5 | CP |
| 0 | CP |
| 3 | CP |
| 0 | CP |
| 4 | CP |
| 3 | CP |
| 0 | depression |
| 0 | depression |
| 10 | depression |
| 0 | depression |
| 0 | depression |
| 0 | depression |
| 5 | depression |
| 4 | depression |
| 0 | depression |
| 0 | depression |
| 3 | depression |
| 0 | depression |
| 2 | depression |
| 6 | depression |
| 0 | depression |

* Two-tailed *t*-tests with unequal variances between each group pair with Holm-Bonferroni method to adjust for multiple comparisons yielded a significance threshold of p=0.0071; this rendered one group difference for CP vs. TD at p=0.006 (2{0-4} vs. 0{0-1}, median {25%-75%}, respectively). No other group differences were detected (p values ranged 0.09 - 0.51).

**Table S2. Heart rate variability measures deployed in the study.**

| Domain | HRV measure | Relation to physiology and references |
|---|---|---|
| *Linear HRV measures* | | |
| Time domain | SDNN (the standard deviation of the RR intervals), RMSSD (the square root of the mean of the squares of the successive differences between adjacent RR intervals); | Long-term fluctuations of R-R intervals (vagal and sympathetic influences) Short-term fluctuations (vagal influences) (Electrophysiology, Task Force of the European Society of Cardiology & the North American Society of Pacing, 1996) (Sassi et al., 2015) |
| Frequency domain | LF (low frequency) power HF (high frequency) power defined as LF, the area (power) in the low frequency band (0.04 to 0.15Hz), and HF, the area (power) in the high frequency band (0.15 to 0.50Hz); | Similar to SDNN Similar to RMSSD (Electrophysiology, Task Force of the European Society of Cardiology & the North American Society of Pacing, 1996) |
| *Nonlinear HRV measures* | | |
| Complexity domain | Sample entropy; Fuzzy entropy | Inflammation, hypoxia (Richman, Lake, & Moorman, 2004) (de Luca & Termini, 1993) |
| Statistical domain | Skewness and asymmetry indices. Moment coefficient of skewness, mode skewness and median skewness. | Inflammation; vagal denervation Asymmetry index was derived as skewness/standard error (Doane & Seward, 2011); ("IBM Knowledge Center," n.d.); (Herry et al., 2016). |
| Geometric domain | Poincaré plots indices SD1 and SD2. Recurrence quantification analysis | Homeokinetic response to surgery stress Cardiac performance |

| | | (Khandoker, Karmakar, Brennan, Palaniswami, & Voss, 2013; Hsu et al., 2012; Herry et al., 2019) |
| --- | --- | --- |
| | | (RQA; Webber & Marwan, 2014) was computed using the following MATLAB code source (Gaoxiang Ouyang Gaoxiang Ouyang, n.d.). Following RQA indices were determined: RR, DET, ENTR and L.(Ouyang, Li, Dang, & Richards, 2008; Ouyang, Zhu, Ju, & Liu, 2014; Frasch et al., 2018) |
| | Phase rectified signal averaging (PRSA). Details in text. | Cardiac performance, maternal – fetal stress (Bauer, Kantelhardt, Barthel, et al., 2006; Bauer, Kantelhardt, Bunde, et al., 2006, Lobmaier et al. 2019; Frasch et al. 2018; Lobmaier et al. 2017) |

**Panel 1**

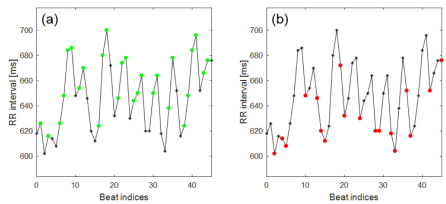

**Panel 2**

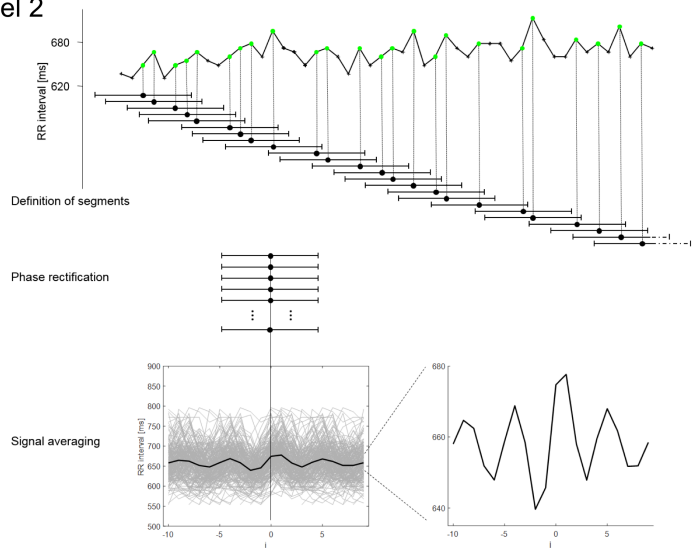

Definition of segments

Phase rectification

Signal averaging

**Panel 3**

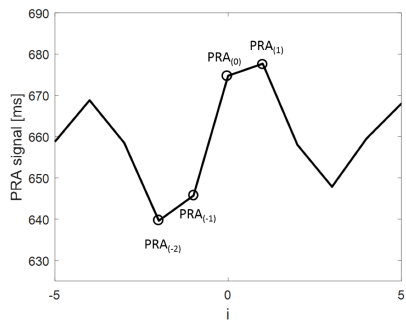

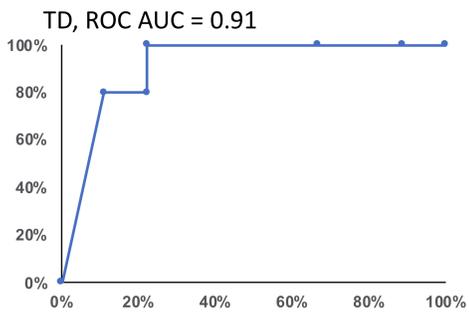

TD, ROC AUC = 0.91

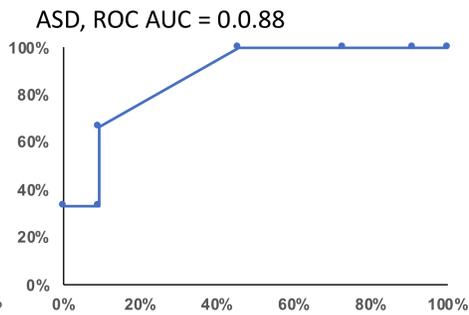

ASD, ROC AUC = 0.0.88

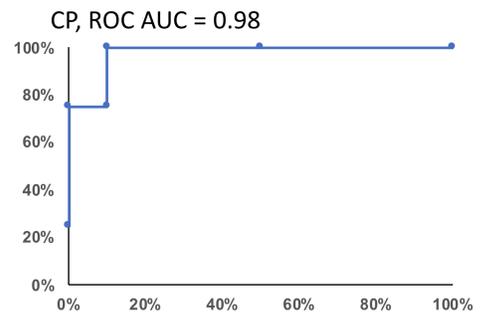

CP, ROC AUC = 0.98

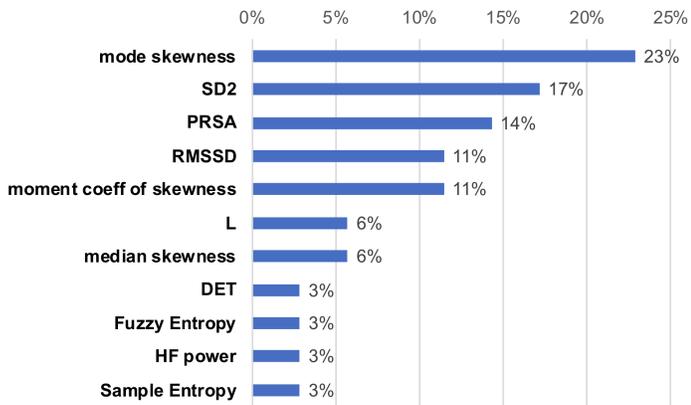

HRV measure importance

| | |
|---|---|
| mode skewness | 23% |
| SD2 | 17% |
| PRSA | 14% |
| RMSSD | 11% |
| moment coeff of skewness | 11% |
| L | 6% |
| median skewness | 6% |
| DET | 3% |
| Fuzzy Entropy | 3% |
| HF power | 3% |
| Sample Entropy | 3% |

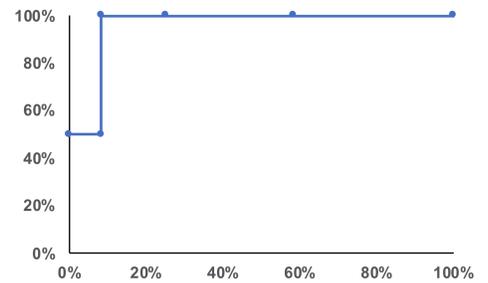

Depression, ROC AUC = 0.96